\documentstyle[aps,prl,twocolumn,epsf,floats]{revtex}
\draft
\begin{document}
\wideabs{

\title{Skyrmions in integral and fractional quantum Hall systems}

\author{
   Arkadiusz W\'ojs}
\address{
   Department of Physics, 
   University of Tennessee, Knoxville, Tennessee 37996, \\
   and Institute of Physics, 
   Wroclaw University of Technology, Wroclaw 50-370, Poland}

\author{
   John J. Quinn}
\address{
   Department of Physics, 
   University of Tennessee, Knoxville, Tennessee 37996}

\maketitle

\begin{abstract}
Numerical results are presented for the spin excitations 
of a two-dimensional electron gas confined to a quantum 
well of width $w$.
Spin waves and charged skyrmion excitations are studied 
for filling factors $\nu=1$, 3, and ${1\over3}$.
Phase diagrams for the occurrence of skyrmions of different 
size as a function of $w$ and the Zeeman energy are calculated.
For $\nu=3$, skyrmions occur only if $w$ is larger than about 
twice the magnetic length.
A general necessary condition on the interaction pseudopotential 
for the occurrence of stable skyrmion states is proposed.
\end{abstract}
}

\paragraph*{Introduction.}
In a numerical study of reversed spin excitations of the spin 
polarized fractional quantum Hall (FQH) state near filling factor 
$\nu={1\over3}$ \cite{Laughlin83}, Rezayi\cite{Rezayi87} found 
that for charged excitations the lowest lying band of states had 
total angular momentum $L$ equal to the total spin $S$ when the 
Zeeman energy $E_{\rm Z}$ was taken to be zero.
In this case the minimum energy occurred at $L=0$ and corresponded 
to a ``spin texture'' containing $K={1\over2}N$ spin flips for 
an $N$ electron system.
Sondhi et al.\cite{Sondhi93} and others\cite{skyrmions_iqhe} 
investigated the $\nu=1$ state and found that for $E_{\rm Z}$ less 
than a critical value $\tilde{E}_{\rm Z}$, the lowest charged 
excitation was a skyrmion\cite{Skyrme61} containing a number $K$ 
of reversed spins that increased as $E_{\rm Z}$ decreased from 
$\tilde{E}_{\rm Z}$ to zero.
Skyrmions have been observed both in magnetization and transport
studies for the $\nu=1$ state\cite{skyrmions_exp} and, when 
$E_{\rm Z}$ was sufficiently decreased by application of hydrostatic 
pressure, for the $\nu={1\over3}$ state\cite{Leadley97}.
The form of the Coulomb pseudopotential in higher Landau levels 
(LL's) suggested that skyrmions would not occur at $\nu=3$, 5, 
\dots\cite{Jain94}.
However, when allowance was made for the softening of the 
pseudopotential associated with finite well width $w$, skyrmions
at $\nu=3$ were predicted\cite{Cooper97} and observed\cite{Song99} 
in sufficiently wide quantum wells.

In this paper we demonstrate the similarities between electron 
and composite Fermion\cite{Jain89} (CF) spin excitations in the 
integral and fractional quantum Hall systems.
We present phase diagrams (in the $w$--$E_{\rm Z}$ plane) 
for the number of spin flips $K$ in the lowest energy charged 
excitation of the $\nu=1$, 3, and ${1\over3}$ fillings.
In addition, we propose necessary conditions on the 
pseudopotentials (applicable both to integral and fractional
filling) for low energy skyrmions at $E_{\rm Z}=0$.

\paragraph*{Model.}
We perform numerical calculations for a system of $N$ electrons
confined to a spherical surface\cite{Haldane83} of radius $R$.
The radial magnetic field is produced by a Dirac monopole at the
center, whose strength $2Q$ is given in units of the quantum of flux, 
$\phi_0=hc/e$, so that $4\pi R^2B=2Q\phi_0$.
The single particle states $\left|Q,l,m\right>$, called monopole 
harmonics\cite{sphere,pseudo}, are eigenfunctions of the orbital 
angular momentum $l$ and its $z$-component $m$.
They form degenerate LL's labeled by $n=l-Q$.
The cyclotron energy $\hbar\omega_c\propto B$ is assumed to be 
much larger than the Coulomb energy scale $E_{\rm C}=e^2/\lambda
\propto\sqrt{B}$ (where $\lambda^2=\hbar c/eB$ is the square of 
the magnetic length).
However, the ratio $\eta=E_{\rm Z}/E_{\rm C}$ is taken as an 
arbitrary parameter.
Finite well width enters the problem only through modifying the 
quasi-2D interaction by replacing $e^2/r$ (where $r$ is the 
in-plane separation) by $V_\xi(r)=e^2\int dz\,dz'\,\xi^2(z)
\xi^2(z')/\sqrt{r^2+(z-z')^2}$.
Here $\xi(z)$ is the envelope function for the lowest subband of
the quantum well.
This change modifies the Coulomb pseudopotential\cite{pseudo} 
$V^{(n)}({\cal R})$, defined as the interaction energy of a pair 
of electrons in the $n$th LL as a function of their relative 
angular momentum ${\cal R}$.
There are four conserved quantum numbers: $L$, the total orbital 
angular momentum, $S$, the total spin, and their projections $L_z$
and $S_z$.
The eigenvalues depend only on $L$ and $S$, and they are therefore
$(2L+1)(2S+1)$ fold degenerate.

\begin{figure}[t]
\epsfxsize=3.40in
\epsffile{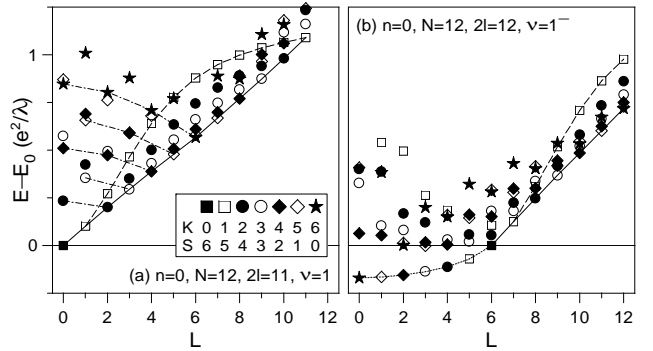}
\caption{
   The energy spectra of 12 electrons in the lowest LL 
   calculated on Haldane sphere with $2l=11$ (a) and 12 (b).}
\label{fig1}
\end{figure}

\paragraph*{Integral filling.}
In Fig.~\ref{fig1}(a) and (b) we present the low energy spectra 
of the $\nu=1$ and $\nu=1^-$ (a single hole in $\nu=1$) states, 
respectively.
In this and all other spectra, only the lowest state at each $L$
and $S$ is shown, $E_0$ is the energy of the lowest maximally 
polarized state ($K=0$), and the Zeeman energy $E_{\rm Z}$ is omitted.

The ferromagnetic ground state of Fig.~\ref{fig1}(a) at $L=0$ and 
$S={1\over2}N=6$ results from the Coulomb interaction even when 
$E_{\rm Z}=0$.
States with different values of $S$ are indicated by the different 
symbols shown in the inset.
The lowest excited state is a spin wave\cite{Kallin84} (SW) 
consisting of a hole in the spin-$\downarrow$ level and an 
electron in the spin-$\uparrow$ level with $L=S=1$.
A dashed line marks the entire single SW band having at $1\le L\le11$ 
(resulting from $\vec{L}=\vec{l}_e+\vec{l}_h$ with $l_e=l_h=l={11\over2}$).
The lowest energy excitation for a given value of either $L$ or $K$ 
occurs at $L=K$ where $K={1\over2}N-S$ is the number of spin flips 
away from the fully polarized ground state.
The (near) linearity of $E(K)$ for this band of states (denoted
by $W_K$) suggests that it consists of $K$ SW's, each with $L=1$, 
which are (nearly) noninteracting.
As shown with the dot-dash lines connecting different states of the 
same number $K$ of $L=1$ SW's, only the $L=K$ state (in which the 
SW's have parallel angular momenta) is noninteracting, and all others 
(at $L<K$) are repulsive.

We have compared the linear $W_K$ energy bands calculated for 
different electron numbers $N\le14$, and found that they all 
have the same slope $u\approx1.15\,e^2/\lambda$ when plotted 
as a function of the ``relative'' spin polarization $\zeta=K/N$.
The fact that $E-E_0=u\zeta$ for the $W_K$ band for every value 
of $N$ has two noteworthy consequences in the 
$N\rightarrow\infty$ limit:
(i) 
For any value of $E_{\rm Z}\ne0$, the interaction energy of each
$W_K$ state, $E-E_0\propto K/N$, is negligible compared to its 
total Zeeman energy, $KE_{\rm Z}$.
(ii) 
The gap for spin excitations at $\nu=1$ equals $E_{\rm Z}$; 
if this gap can be closed (e.g., by applying hydrostatic pressure), 
the $\nu=1$ ferromagnet becomes gapless and the density of states 
for the $W_K$ excitations becomes continuous.

Because of the exact particle--hole symmetry in the lowest LL,
the $\nu=1^-$ state whose spectrum appears in Fig.~\ref{fig1}(b)
can be viewed as containing either one hole or one reversed spin 
electron in a $\nu=1$ ground state.
The band of states with $0\le L\le5$ and $S=L$ (dotted line) is 
the skyrmion band denoted by $S_K$.
Its energy increases monotonically with $S$ and $L$.
For $6\le L\le12$ the single SW band (dashed line) and band of 
$K$ SW's each with $L=1$ (solid line) resemble similar bands in 
Fig.~\ref{fig1}(a), except that their angular momenta are added 
to that of the hole which has $l_h=l=6$.

Fig.~\ref{fig1} completely ignores the Zeeman energy.
The total Zeeman energy measured from the fully polarized state
is proportional to $K$.
The total energy of the skyrmion band is $E(K)=E_S(K)+KE_{\rm Z}$
and the lowest $S_K$ state occurs when $E(K)$ has its minimum.
If we very roughly approximate the skyrmion energy in a finite
system by $E_S(K)\approx E_S({1\over2}N)+\beta S^2$, where $\beta\ge0$ 
is a constant, this minimum occurs at $K={1\over2}(N-E_{\rm Z}/\beta)$ 
spin flips.
This vanishes when $E_{\rm Z}=\beta N$, defining the critical 
value, $\tilde{E}_{\rm Z}$, and it reaches its maximum value 
($K={1\over2}N$ or complete depolarization) when $E_{\rm Z}=0$.
At such $E_{\rm Z}$ that the ground state at $\nu=1^\pm$ is a finite 
size skyrmion, its gap for spin excitations is much smaller than (and 
largely independent of) $E_{\rm Z}$.
This is in contrast to the exact $\nu=1$ filling and allows spin 
coupling of the electron system to the magnetic ions, nuclei, or 
charged excitons.

\begin{figure}[t]
\epsfxsize=3.40in
\epsffile{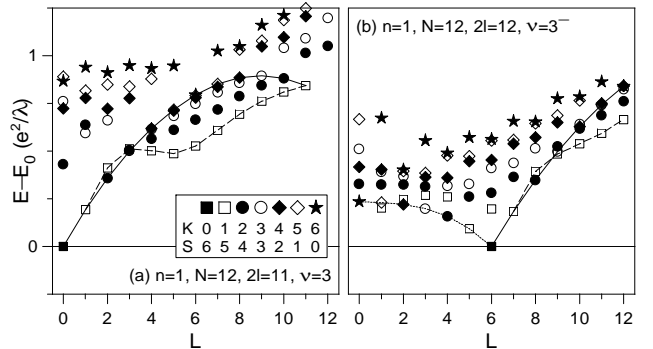}
\caption{
   Same as Fig.~\ref{fig1} but for the first excited LL.}
\label{fig2}
\end{figure}

In Fig.~\ref{fig2} we show the numerical results analogous to 
those in Fig.~\ref{fig1} but for the $n=1$ LL.
Features clearly apparent in the lowest LL are now absent.
For example, the $W_K$ band in Fig.~\ref{fig2}(a) departs noticeably 
from linearity, and it does not generally lie below the the single 
SW band (dashed line).
More striking is the fact that the $S_K$ band of Fig.~\ref{fig2}(b)
goes above the single hole state at $L=6$, in contrast to the behavior 
in Fig.~\ref{fig1}(b).
Therefore skyrmions are not the lowest energy charged excitations 
in excited LL's even when $E_{\rm Z}=0$.
This effect was first predicted by Jain and Wu\cite{Jain94}.

The only difference between the filling factors $\nu=3$, 5, \dots\ 
and $\nu=1$ is that the monopole harmonics $\left|Q,l=Q+n,m\right>$ 
correspond to the excited LL instead of the lowest.
Matrix elements of the Coulomb interaction $e^2/r$ between these 
higher monopole harmonics give a different pseudopotential 
$V^{(n)}({\cal R})$ from that for $n=0$.
Though one might expect skyrmions to be the lowest energy charged 
excitations in this case, the change in the pseudopotential from 
$V^{(0)}({\cal R})$ to $V^{(n)}({\cal R})$ with $n\ge1$ causes the 
charged spin flip excitations to have higher energy than the single 
hole or reversed spin electron.

\begin{figure}[t]
\epsfxsize=3.40in
\epsffile{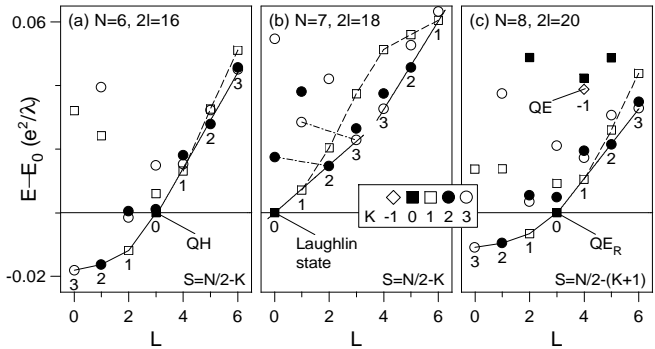}
\caption{
   The energy spectra of $N=6$ to 8 electrons calculated 
   on Haldane sphere at the values of $2l$ corresponding 
   to $\nu={1\over3}^-$ (a), $\nu={1\over3}$ (b), and 
   $\nu={1\over3}^+$ (c).}
\label{fig3}
\end{figure}

\paragraph*{Fractional filling.}
Since the CF picture\cite{Jain89} describes the FQH effect in terms 
of integral filling of effective CF levels, it is interesting to ask
\cite{skyrmions_fqhe} if spin excitations analogous to the SW's and 
skyrmions occur at Laughlin fractional fillings $\nu=(2p+1)^{-1}$ 
(where $p=1$, 2, \dots).
In Fig.~\ref{fig3} we display numerical results for $\nu\approx{1\over3}$.
The values of $N$ and $2l$ in frames (b), (a), and (c) correspond 
to a Laughlin $\nu\approx{1\over3}$ condensed state, Laughlin 
quasihole (QH), and Laughlin quasielectron (QE) or reversed spin 
quasielectron (QE$_{\rm R}$), respectively.
For each of these cases the lowest CF LL has a degeneracy of seven.
Clearly the single SW dispersion (dashed line) and the linear $W_K$ 
band (solid line) both appear in Fig.~\ref{fig3}(b).
The $S_K$ bands beginning at $L=0$ lie below the single QH 
state (a) and below the single QE$_{\rm R}$ state (c).
The solid and dashed lines at $3\le L\le6$ in Fig.~\ref{fig3}(a)
and (c) are completely analogous to those in Fig.~\ref{fig1}(b),
and correspond to the single SW band and the $W_K$ band, except 
that their angular momenta are added to $l_{\rm QH}=3$ or 
$l_{{\rm QE}_{\rm R}}\!=3$.
What is clearly different from the $\nu=1$ case is the smaller
energy scale, and a noticeable difference between the $\nu=
{1\over3}^-$ (QH) and $\nu={1\over3}^+$ (QE$_{\rm R}$) spectra.
Since the QH--QH and QE$_{\rm R}$--QE$_{\rm R}$ interactions 
are known to be different \cite{qer}, this lack of 
QH--QE$_{\rm R}$ symmetry is not unexpected.
It implies a lack of symmetry between the CF skyrmion (QE$_{\rm R}
+K$ SW) and CF antiskyrmion (QH$+K$ SW) states in contrast to the 
skyrmion--antiskyrmion symmetry of $\nu=1$.
Because the CF skyrmion energy scale is so much smaller than 
$E_{\rm C}$ at $\nu=1$, the critical $E_{\rm Z}$ at which skyrmions 
are stable is correspondingly smaller\cite{Leadley97}.

\begin{figure}[t]
\epsfxsize=3.40in
\epsffile{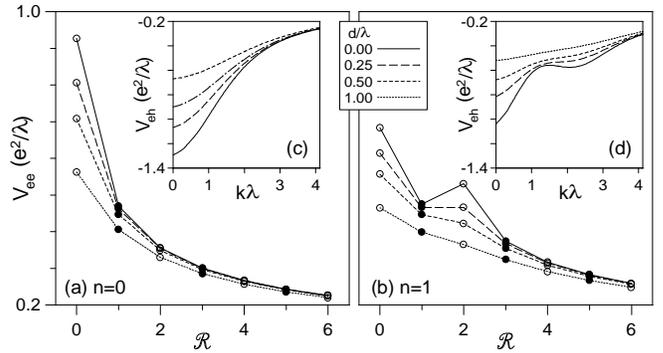}
\caption{
   The $e$--$e$ interaction pseudopotentials in the lowest (a) and 
   first excited (b) LL, calculated for the potential $V_d(r)$.
   Only data for ${\cal R}\le6$ is shown, and open and closed circles 
   distinguish between singlet and triplet states.
   Insets: corresponding $e$--$h$ pseudopotentials; $k$ is the 
   $e$--$h$ wave vector.}
\label{fig4}
\end{figure}

\paragraph*{Effect of finite well width.}
As suggested by Cooper\cite{Cooper97} and confirmed experimentally 
by Song et al.\cite{Song99}, skyrmions become the lowest energy 
charged excitations in higher LL's if the quantum well is 
sufficiently wide.
The finite well width $w$ can be accounted for by using effective
potential $V_\xi(r)$ and selecting a subband wave function $\xi(z)$ 
appropriate to the depth and width of the quantum well.
In Fig.~\ref{fig4} we show the $e$--$e$ and $e$--$h$ pseudopotentials 
for the lowest (ac) and excited (bd) LL as a function of a parameter 
$d$ which is proportional to $w$.
In the calculation we have used a potential $V_d(r)=e^2/
\sqrt{r^2+d^2}$ as done earlier by He et al.\cite{He90}.
Comparing the resulting pseudopotentials with those obtained
using an envelope function $\xi_0(z)\propto\cos(\pi z/w)$
appropriate to the lowest subband of an infinitely deep 
quantum well gives $w\approx5d$ ($w$ is slightly larger than 
the actual width of a finite depth well).
It is clear that finite width must have the largest effect on 
those pseudopotential coefficients corresponding to the smallest 
average $e$--$e$ or $e$--$h$ separation.
As a result, increasing $w$ causes suppression of the maxima
of $V_{\rm ee}({\cal R})$ and of the minima of $V_{\rm eh}(k)$
characteristic of the excited LL's.
This makes the $n=1$ pseudopotentials (and, in consequence, the
many-body spectra of Fig.~\ref{fig2}) more similar to those of 
the lowest LL.

\begin{figure}[t]
\epsfxsize=3.40in
\epsffile{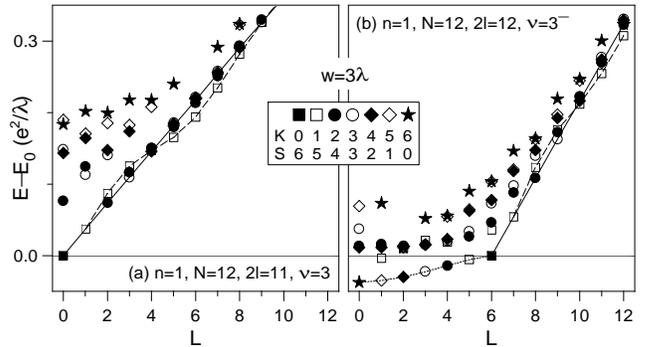}
\caption{
   Same as Fig.~\ref{fig2} but for finite well width $w=3\lambda$.}
\label{fig5}
\end{figure}

In Fig.~\ref{fig5} we show the same energy spectra as given 
in Fig.~\ref{fig2}, but for the Coulomb pseudopotential in the
$n=1$ LL replaced by one appropriate for $w=3\lambda$.
The $W_K$ band in Fig.~\ref{fig5}(a) is now much closer to linear 
with $K$, and the $S_K$ band in Fig.~\ref{fig5}(b) now has $E<E_0$.
We have done similar calculations for the $n=2$ LL with similar 
results.
These results show that skyrmions are the lowest excitations in higher 
LL's if $w$ is sufficiently large.

\begin{figure}[t]
\epsfxsize=3.40in
\epsffile{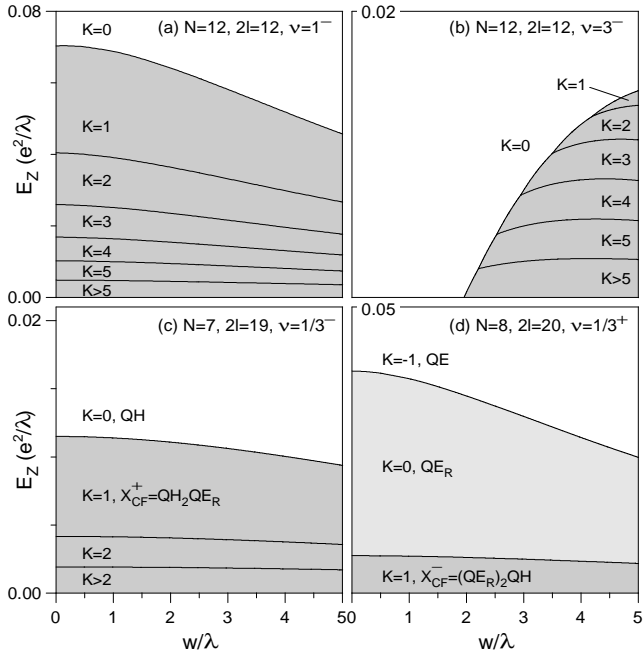}
\caption{
   Phase diagrams for the occurrence of skyrmions with different 
   $K$ as a function of well width $w$ and Zeeman energy $E_{\rm Z}$,
   calculated at $\nu=1^-$ (a), $3^-$ (b), ${1\over3}^-$ (c), and
   ${1\over3}^+$ (d).
   Numbers in top-left corners give the upper bounds of the vertical 
   axes (the lower bound are zero in all frames).}
\label{fig6}
\end{figure}

In Fig.~\ref{fig6} we sketch the phase diagrams (in the 
$w$--$E_{\rm Z}$ plane) for charged excitations at the 
integral filling of the lowest and excited LL's ($\nu=1$ and 3), 
and at the fractional filling $\nu={1\over3}$.
For $n=0$ ($\nu=1$ or ${1\over3}$), skyrmions are the lowest charged 
excitations below a critical value of $E_{\rm Z}$ which is relatively 
insensitive to the width $w$.
As $E_{\rm Z}$ is decreased, larger skyrmions (with increasing $K$) 
become the lowest energy states.
For $n=1$, no skyrmions occur unless $w\le2\lambda$ (we have checked 
that this value remains correct for small skyrmions in the 
$N\rightarrow\infty$ limit).

\paragraph*{Pseudopotentials and skyrmion stability.}
Only the leading pseudopotential coefficients $V(0)$, $V(1)$, $V(2)$, 
corresponding to small average in-plane $e$--$e$ separations, are 
strongly influenced by finite $w$.
In fact, the change in the energy $E_S(K)-E_0$ of the skyrmion band 
from positive to negative occurred in the $n=1$ LL only when the 
pseudopotential coefficient $V(2)$ was quite strongly affected by 
the increase in $w$.
For this reason, we investigate $E_S(K)$ for a simple model 
pseudopotential with the following properties: 
(i) $V(0)$ is sufficiently large to cause decoupling of the many
body states that avoid all ${\cal R}=0$ pairs (i.e., the skyrmion 
states) from all other states that contain some ${\cal R}=0$ pairs;
(ii) behavior of $V({\cal R})$ between ${\cal R}=1$ and 3 can be
varied similarly to how $V^{(1)}$ varies with increasing $w$.
We choose the simplest possible model pseudopotential with these
properties by defining $U_x({\cal R})$ as follows: $U_x(0)=\infty$, 
$U_x(1)=1$, $U_x(2)=x$, and $U_x({\cal R})=0$ for ${\cal R}>2$.
This choice of $U_x$ guarantees that skyrmions are its only finite 
energy eigenstates, and their energy depends on one free parameter 
$x$.
A simple relation between the fractional grandparentage coefficients
\cite{pseudo} ${\cal G}_K({\cal R})$ at ${\cal R}=0$, 1, and 2 yields 
$E_S(K)-E_0={\cal G}_K(2)(x-\alpha)$, where ${\cal G}_K(2)$ is a 
positive constant and $\alpha^{-1}=2-(N-1)^{-1}$.
Since for every value of $K$, $E_S(K)-E_0$ changes sign at 
$x=\alpha$, and $\alpha\rightarrow{1\over2}$ in large systems, 
we conclude that skyrmions are the lowest charged excitations 
when $U_x(2)$ drops below half of $U_x(1)$, and $U_x$ becomes 
superlinear between ${\cal R}=1$ and 3.
Therefore we suggest that for both integral and fractional quantum
Hall states the stability of skyrmion states requires an effective
pseudopotential for which 
(i) $V(0)$ is large enough to cause Laughlin correlations, and
(ii) $V(2)$ is less than or equal to half of $V(1)+V(3)$.

\paragraph*{Acknowledgment.}
The authors acknowledge partial support by the Materials Research 
Program of Basic Energy Sciences, US Department of Energy, and thank
P. Hawrylak, M. Potemski, and I. Szlufarska for helpful discussions.
JJQ thanks Natl.\ Magnet.\ Lab., Tallahassee, and UNSW, Sydney, for 
their hospitality.
AW acknowledges support from KBN grant 2P03B11118.


\begin{references}

\bibitem{Laughlin83}
R. B. Laughlin,
   Phys. Rev. Lett. {\bf50} (1983) 1395.

\bibitem{Rezayi87}
E. H. Rezayi, 
   Phys. Rev. B {\bf36} (1987) 5454.

\bibitem{Sondhi93}
S. L. Sondhi, A. Karlhede, S. A. Kivelson, and E. H. Rezayi,
   Phys. Rev. B {\bf47} (1993) 16419.

\bibitem{skyrmions_iqhe}
L. Brey, H. A. Fertig, R. C\^ot\'e, and A. H. MacDonald,
   Phys. Rev. Lett. {\bf75} (1995) 2562;
A. H. MacDonald, H. A. Fertig, and L. Brey, 
   {\sl ibid.} {\bf76} (1996) 2153;
H. A. Fertig, L. Brey, R. C\^ot\'e, and A. H. MacDonald,
   Phys. Rev. B {\bf50} (1994) 11018;
X. C. Xie and S. He, 
   {\sl ibid.} {\bf53} (1996) 1046;
J. J. Palacios, D. Yoshioka, and A. H. MacDonald,
   Phys. Rev. B {\bf54} (1996) 2296;
H. A. Fertig, L. Brey, R. C\^ot\'e, A. H. MacDonald, 
A. Karlhede, and S. L. Sondhi,
   {\sl ibid.} {\bf55} (1997) 10671.

\bibitem{Skyrme61}
T. Skyrme,
   Proc. R. Soc. London A {\bf262} (1961) 237.

\bibitem{skyrmions_exp}
S. E. Barrett, G. Dabbagh, L. N. Pfeiffer, K. W. West, and R. Tycko,
   Phys. Rev. Lett. {\bf74} (1995) 5112;
D. K. Maude, M. Potemski, J. C. Portal, M. Henini, L. Eaves, 
G. Hill, and M. A. Pate, 
   {\sl ibid.} {\bf77} (1996) 4604;
E. H. Aifer, B. B. Goldberg, and D. A. Broido,
   {\sl ibid.} {\bf76} (1996) 680;
R. Tycko, S. E. Barrett, G. Dabbagh, L. N. Pfeiffer, and K. W. West, 
   Science {\bf268} (1995) 1460.

\bibitem{Leadley97}
D. R. Leadley, R. J. Nicholas, D. K. Maude, A. N. Utjuzh, 
J. C. Portal, J. J. Harris, and C. T. Foxon,
   Phys. Rev. Lett. {\bf79} (1997) 4246.


\bibitem{Jain94}
J. K. Jain and X. G. Wu,
   Phys. Rev. B {\bf49} (1994) 5085;
X. G. Wu and S. L. Sondhi,
   {\sl ibid.} {\bf51} (1995) 14725.

\bibitem{Cooper97}
N. R. Cooper,
   Phys. Rev. B {\bf55} (1997) R1934.

\bibitem{Song99}
Y. Q. Song, B. M. Goodson, K. Maranowski, and A. C. Gossard,
   Phys. Rev. Lett. {\bf82} (1999) 2768.

\bibitem{Jain89}
J. K. Jain,
   Phys. Rev. Lett. {\bf63} (1989) 199.

\bibitem{Haldane83}
F. D. M. Haldane,
   Phys. Rev. Lett. {\bf51} (1983) 605.

\bibitem{sphere}
T. T. Wu and C. N. Yang,
   Nucl. Phys. B {\bf107} (1976) 365.

\bibitem{pseudo}
A. W\'ojs and J. J. Quinn,
   Acta Phys. Pol. A {\bf96} (1999) 403;
   Philos. Mag. B {\bf80} (2000) 1405;
J. J. Quinn and A. W\'ojs,
   J. Phys.: Cond. Mat. {\bf12} (2000) R265.

\bibitem{Kallin84}
C. Kallin and B. I. Halperin,
   Phys. Rev. B {\bf30} (1984) 5655.

\bibitem{skyrmions_fqhe}
R. K. Kamilla, X. G. Wu, and J. K. Jain,
   Solid State Commun. {\bf99} (1996) 289;
A. H. MacDonald and J. J. Palacios
   Phys. Rev. B {\bf58} (1998) R10171.

\bibitem{qer}
I. Szlufarska, A. W\'ojs, and J. J. Quinn,
   Phys. Rev. B {\bf64} (to appear October 15, 2001).

\bibitem{He90}
S. He, F. C. Zhang, X. C. Xie, and S. Das Sarma, 
   Phys. Rev. B {\bf42} (1990) R11376.

\end{references}
\end{document}